\documentclass[preprint,showpacs]{revtex4}

\usepackage{times}
\usepackage{bm}
\usepackage[dvips]{graphicx}
\usepackage{amsmath}
\usepackage{bm}
\usepackage{natbib}
\usepackage{color}
\newcommand{\ket}[1]{|#1\rangle}
\newcommand{\bra}[1]{\langle #1|}

\begin{document}
\title{Splitting of the Zero-Energy Landau Level and Universal Dissipative Conductivity at Critical Points in Disordered Graphene}
\author{Frank Ortmann,$^{1}$ and Stephan  Roche,$^{1,2}$}
\affiliation{
$^1$CIN2 (ICN-CSIC) and Universitat Aut\'{o}noma de Barcelona, Catalan Institute
 of Nanotechnology, Campus UAB, 08193 Bellaterra, Spain\\ 
$^2$ICREA, Instituci\'{o} Catalana de Recerca i Estudis Avan\c{c}ats, 08070 Barcelona, Spain}
\date{\today}

\begin{abstract}
We report on robust features of the longitudinal conductivity ($\sigma_{xx}$) of the graphene zero-energy Landau level in presence of disorder and varying magnetic fields. By mixing an Anderson disorder potential with a low density of sublattice impurities, the transition from metallic to insulating states is theoretically explored as a function of Landau-level splitting, using highly efficient real-space methods to compute the Kubo conductivities (both $\sigma_{xx}$ and Hall $\sigma_{xy}$). As long as valley-degeneracy is maintained, the obtained critical conductivity $\sigma_{xx}\simeq 1.4 e^{2}/h$ is robust upon disorder increase (by almost one order of magnitude) and magnetic fields ranging from about 2 to 200 Tesla. When the sublattice symmetry is broken, $\sigma_{xx}$ eventually vanishes at the Dirac point owing to localization effects, whereas the critical conductivities of pseudospin-split states (dictating the width of a $\sigma_{xy}=0$ plateau) change to $\sigma_{xx}\simeq e^{2}/h$, regardless of the splitting strength, superimposed disorder, or magnetic strength.  These findings point towards the non dissipative nature of the quantum Hall effect in disordered graphene in presence of Landau level splitting.
\end{abstract} 

\maketitle


\textit{Introduction}.-The massless Dirac fermion nature of low-energy excitations in monolayer graphene remarkably manifests in the high magnetic field regime, where the energy spectrum splits up into non-equidistant Landau levels (LL) given by $E_{n}=sgn(N)\sqrt{2\hbar{v_{F}}^{2}eB|N|}$ \cite{McClure,Castro2009}.  One fundamental signature of such peculiar spectrum is the existence of a fourfold degenerate zero-energy LL (twofold valley and spin degeneracies) where electrons and holes coexist.  As a result, the integer quantum Hall effect (QHE) \cite{Klitzing1980} (measured in conventional two-dimensional electron gas) transforms to a half-integer (anomalous) QHE in graphene, with a quantized Hall conductivity given by $\sigma_{xy}= 4e^{2}/h\times(N+1/2)$ \cite{Novoselov2005,Zhang2005,Goerbig2011}.  

Such anomalous QHE is tightly interwoven with the $\pi$-Berry phase and pseudospin degree of freedom, and occurs as long as $K$ and $K'$ valleys remain decoupled \cite{Ostrovsky2008}.  In contrast, if disorder breaks sublattice symmetry and strongly mixes valleys, the QHE in disordered graphene is predicted not to differ from other two-dimensional systems, recovering $\sigma_{xy}= 2Ne^{2 }/h$, with $N$ an integer \cite{Ostrovsky2008,Aleiner2006,Altland2006}.     Several experiments performed in high-mobility samples have revealed an additional quantized Hall plateau at $\sigma_{xy}=0$, evidencing a splitting of the zero-energy LL which could result from spin and/or sublattice-degeneracy lifting, stemming respectively from Zeeman interaction, sublattice symmetry-breaking mechanisms \cite{Li2009}, or electron-electron interactions \cite{Zhang2006,Jian2007,Nomura2006,Young2012}. 

The origin of such quantized plateau at $\sigma_{xy}=0$ has been further discussed in relation with the measurement of a finite value of $\sigma_{xx}$ at the Dirac point, suggesting an unconventional dissipative nature of the QHE, however difficult to decipher \cite{Abanin2006,Abanin2007,Zhang2010,Jia2008,Checkelsky2008}. Indeed, a theoretical scenario proposes the existence of a dissipative QHE phenomenon near the Dirac point (with finite conductivity $\sigma_{xx}\sim 1-2 e^{2}/h$) which would be conveyed by counter propagating (gapless) edge states carrying opposite spin \cite{Abanin2006,Abanin2007,Zhang2010}. Finite $\sigma_{xx}$ ($\simeq e^{2}/(\pi h)$) at the Dirac point has been also obtained from numerical simulations in tight-binding models of disordered graphene (introducing either bond disorder \cite{Jia2008} or random magnetic flux \cite{Schweitzer2008}), and related with the formation of extended states centered at zero energy (but in absence of a fully quantized $\sigma_{xy}$ \cite{Jia2008}). 

Differently, other experiments have reported a strongly divergent resistivity at the Dirac point ($\sigma_{xx}\to 0$ in the zero-temperature limit) \cite{Checkelsky2008} which has been analyzed in terms of a Kosterlitz-Thouless metal-insulator transition \cite{Nomura2009}. The observation of a temperature-dependent activated behavior of $\sigma_{xx}(T)$  further points towards a non-dissipative nature of the plateau $\sigma_{xy}=0$ for a spin-splitting gap opening \cite{Giesbers2009,Kurganova2011,Zhao2012}.

This shows that the experimental literature on QHE in graphene is very rich and diversified, mainly because there exist various qualities of graphene material (epitaxial, CVD-grown, or exfoliated from graphite) as well as experimental measurement conditions (silicon oxide or boron-nitiride substrate and suspended graphene…). Disorder can also exist in a large variety of flavors (adsorbed impurities, vacancies, grain boundaries…) which thus demand for transport universalities to be established.

In this Letter, the magnetic field dependent-fingerprints of the dissipative conductivity ($\sigma_{xx}$) of disordered graphene are explored with and without energy level splitting. Using a tight-binding Hamiltonian and real space order-$N$ quantum transport approaches, the Kubo conductivities $\sigma_{xx}$ and $\sigma_{xy}$ are computed as a function of disorder and magnetic field. By tuning the contribution of valley mixing, universal features of $\sigma_{xx}$ are unveiled, such as a robust critical conductivity of $\sigma_{xx}\simeq 1.4 e^{2}/h$ at the Dirac point as long as valley degeneracy is unbroken. In contrast, if sublattice symmetry is lifted by some impurity potential, pseudospin-split states are generated and found to convey different critical bulk conductivities $\sigma_{xx}\simeq e^{2}/h$, regardless the splitting strength and magnitude of the magnetic field. In between pseudospin-split critical states, $\sigma_{xx}$ is found to eventually vanish in the zero temperature limit owing to intervalley-induced localization effects, in conjunction with the appearance of the quantized $\sigma_{xy}=0$ plateau. These findings establish different critical values of the dissipative conductivity at the  center of Landau levels of lowest energies, together with a clarification on the non-dissipative nature of the QHE in disordered graphene in presence of energy level splitting induced by sublattice symmetry breaking. 

\textit{Methodology}- Electronic and transport properties are investigated by using a simple $\pi$-$\pi$* orthogonal tight-binding (TB) model with nearest neighbor hopping $\gamma_0$ (taken as 2.7 eV)
\begin{equation}
{\cal H} =  \sum_{\alpha}V_{\alpha}|\alpha \rangle\langle \alpha
|- \gamma_0 \sum_{\langle\alpha,\beta\rangle } e^{-i\varphi_{\alpha
\beta}} \ket{\alpha}\bra{\beta},
\label{eq:Hamiltonian}
\end{equation}
where the magnetic field is introduced through a Peierls phase~\cite{Peierls} with a magnetic flux per hexagon being equal to  $\phi=\oint{\bf A}\cdot d{\bf l}=h/e\sum_{\rm hexagon}\varphi_{\alpha\beta}$. A suitable gauge is chosen, allowing the calculation of transport properties in disordered graphene with realistic values of $B$ (here varied from about 2 to 200 Tesla). An Anderson disorder is first introduced through a modulation of the potential profile, by taking onsite energies at random within $[-W/2, W/2]$ ($\gamma_{0}$-units) where $W$ gives the disorder strength. This is a commonly used disorder model for exploring the metal-insulator transition in low dimensional systems (with or without applied magnetic field) \cite{Evers2008,Sheng2006}.

Quantum transport in high magnetic fields is studied with order-$N$ computational schemes for $\sigma_{xx}(E,B)$ \cite{Roche1}, as well as for the Hall conductivity $\sigma_{xy}(E,B)$ \cite{Ortmann2012}, using real space implementations of the Kubo approach. The scaling properties of $\sigma_{xx}$ can be followed through the dynamics of electronic wavepackets using \cite{Roche1}
\begin{equation}
\sigma(E,t)=
e^{2}\rho(E) \frac{1}{t}\Delta X^{2}(E,t)
\label{eq:kubo}
\end{equation}
where $\rho(E)$ is the density of states (DOS) and $\Delta X^{2}(E,t)$ is the mean quadratic displacement of the wave packet at energy $E$ and time $t$:
\begin{equation}
\Delta X^{2}(E,t) = \frac{\displaystyle {\large\rm Tr}\bigl[ \delta(E-{\cal H}) | \hat{X}(t)- \hat{X}(0) |^2 \bigr]}
{ \strut\displaystyle {\large\rm Tr}[\delta(E-{\cal H})]}
\label{DeltaX2}
\end{equation}
A key quantity is the diffusion coefficient defined as $D_x(E,t) =\Delta X^{ 2}(E_{F},t)/t$, which gives the conductivity through Eq.~(\ref{eq:kubo}) at a certain timescale. The spin degree of freedom is included as a factor of two for $\sigma$ and $\rho$, while calculations are performed with system sizes containing up to several tens of millions of carbon atoms and energy resolution down to $5\cdot 10^{-5}\gamma_{0}$. All the information about multiple scattering effects is contained in the time-dependence of $D_x(E,t)$.  The trace in Eq.(\ref{DeltaX2}) is evaluated numerically using random-phase wavepackets $|\varphi_\text{RP}\rangle$ according to ${\large\rm Tr}\bigl[ ...\bigr]\to N_s\langle\varphi_\text{RP}|...|\varphi_\text{RP}\rangle$ \cite{Roche1}.  Such method has been now widely used for studying strongly disordered materials \cite{REC}. 

The Hall Kubo conductivity is also computed from the time evolution of random-phase wavepackets $|\varphi_\text{RP}\rangle$ and the Lanczos method, by rewritting $\sigma_{xy}(E)$ as
\begin{equation}
\sigma_{xy}(E)=-\frac{2}{V}\int_{0}^{\infty}dte^{-\eta t/\hbar}\int_{-\infty}^{\infty}dE'f(E'-E) 
\;\text{Re}\left[\langle\varphi_\text{RP}|
\delta(E'-\hat{H})\hat{j}_{y}\frac{1}{E'-\hat{H}+i\eta}\hat{j}_{x}(t)|
\varphi_\text{RP}\rangle\right],
\end{equation}

\noindent
with $\hat{j}_{x}=\frac{ie_0}{\hbar}[H,\hat{X}]$, the current operator ($\hat{X}$ the position operator), while $\eta\to 0$ is a small parameter required for achieving numerical convergence. A new algorithm has been implemented following prior studies \cite{Ortmann2012}. 

\begin{figure}[htbp]
\begin{center}
\leavevmode
\includegraphics[width=82mm]{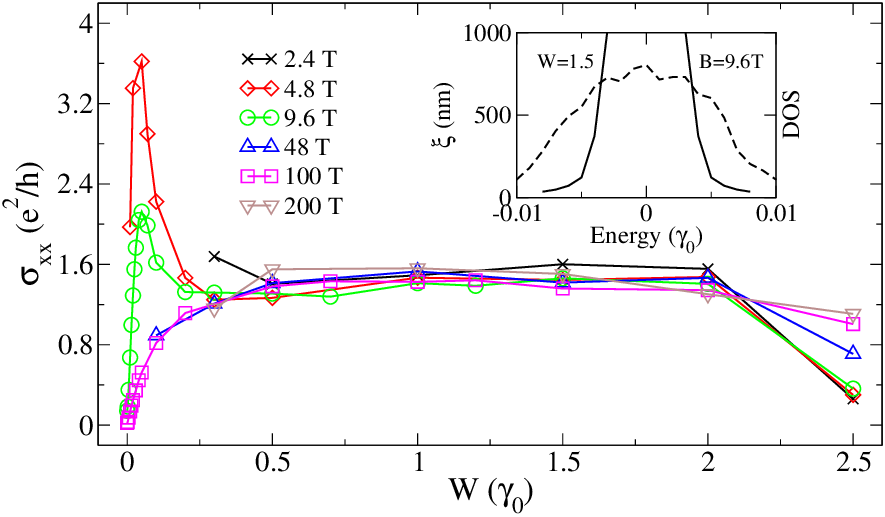}
\caption{(color online) Main frame: Zero-energy conductivity versus disorder strength $W$, and varying magnetic field (from 2.4 to 200 Tesla). Inset: Localization length $\xi(E)$ for the zero-energy Landau level (solid line) and density of states (dashed line, and arbitrary unit) for $W$=1.5 and $B$=9.6 Tesla.}
\label{UniversalSigma}
\end{center}
\end{figure}

\textit{Critical conductivity $\sigma_{xx}(E,B,W)$ of the zero energy LL}.-We study the evolution of $\sigma_{xx}(E,B,W\leq 2.5)$ in presence of Anderson disorder (which preserves chiral symmetry). At the Dirac point, different transport regimes are identified in Fig.\ref{UniversalSigma} (main frame) (for $t=12$ ps, the maximum computed time).  At $W=0$, all states are localized by the magnetic field ($\sigma_{xx}$ tends numerically to zero), but small disorder brings delocalization into play, as manifested by the enhancement of $\sigma_{xx}$ with $W$. For non-zero disorder up to $W$=2,  $D_x(E,t)$ are found to saturate to some maximum values in the long-time limit ($D_x(E,t)\to D_{\text{max}}(E)$), pinpointing the establishment of a diffusive regime and absence of Anderson localization effects. Additionally, for small enough disorder (up to $W\simeq 0.1$), $\sigma_{xx}$ increases roughly linearly with $W$ whatever the strength of the field (tuned from 4.8 to 100 Tesla). The value of $d\sigma_{xx}/dW$ depends on the magnetic field (being larger for lower $B$) as expected from the scaling of the magnetic length $l_B\propto B^{-1/2}$, which suggests reduced disorder-induced delocalization effects as $B$ is increased (and $l_B$ is shortened). 

A remarkable saturation of $\sigma_{xx}$ to a constant value $\simeq 1.4 e^2/h$ is further obtained for a large range of disorder strengths $W\in [0.3,2]$ and magnetic fields varying between 2.4 Tesla and 200 Tesla (up to two orders of magnitudes). This value identifies the critical regime in which the interplay between disorder and magnetic field preserves extended states only at the center of the LL, while remaining states become localized (a key ingredient of the QHE theory). This is further rationalized by analyzing the nature of electronic states in the vicinity of the Dirac point (for disorder $W\leq 2$). By converting the time propagation of wavepackets to their spatial spreading and infering a corresponding length-dependent conductivity $\sigma_{xx}(L)$, the localization lengths are extracted at selected energies by fitting $\sigma_{xx}(L)$ with an exponentially decaying function. The typical behavior of $\xi(E)$ is illustrated for $W=1.5$ in Fig.\ref{UniversalSigma} (inset),  with a diverging $\xi(E)$ with energy lowering. When disorder exceeds $W$=2.5, all states (including the states at the Dirac point) become localized and the system is driven to the insulating state (for experimentally accessible values of $B$) with disappearance of the QHE regime, in agreement with prior numerical studies \cite{Sheng2006}. 

\begin{figure}[htbp]
\begin{center}
\leavevmode
\includegraphics[width=82mm]{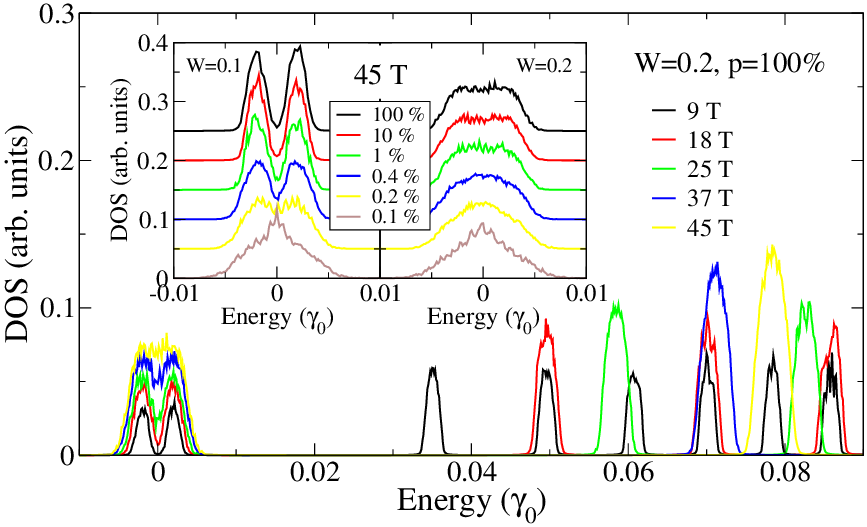}
\caption{(color online) Density of states for various magnetic fields $B$ and disorder strengths $W$ and adding a superimposed A/B-sublattice impurity potential given by $V_\text{A}$ (see main text), which splits the zero-energy Landau level. Main frame: full coverage of sublattice impurities with $pV_\text{A}=V_\text{A}=$0.002. Inset: dilute random symmetry breaking potential with indicated percentage $p$ (constant total strength $pV_\text{A}$=0.002) for two values of $W=0.1$ (left) and $W=0.2$ (right).}
\label{DOS}
\end{center}
\end{figure}

\textit{Critical conductivity of pseudospin split states}.-The robustness of the obtained critical value at the Dirac point is further investigated by adding a density of impurities which break the local A/B sublattice symmetry. To induce pseudospin-splitting, we use a heuristic model which consists in shifting all onsite energies of A (and B) lattice sites by a constant quantity $V_\text{A}$ (and $V_\text{B}$).  We first simplify to the situation where all A and B sites are differenciated in energy according to $V_\text{A}=-V_\text{B}$, which induces a splitting gap of $V_\text{A}-V_\text{B}=2V_\text{A}$. The Anderson disorder potential is maintained but with $|V_\text{A}|\ll W$, potentially masking the formation of a pseudospin-split gap (see uppermost curve in Fig. \ref{DOS} inset). The superposition of both potentials mimics some weak imbalance in the adsorption site in the sense of a slightly preferred sublattice.  We note that recent experiments curiously report such possibility of imbalance doping or structural damage \cite{Zhao2011,Terrones2012,ABSB1,ABSB2}.

Fig. \ref{DOS} (main frame) shows the density of states for $W=0.2$ and $V_\text{A}=0.002$ which corresponds to a weak  imbalance of adsorption on one sublattice. This imbalance splits the zero energy LL for magnetic fields as low as $B=$9 Tesla,  but the splitting is reduced with increasing field, and becomes hardly visible for 45 Tesla  (close-up right part of inset, black). We note that the sequence of higher LL does not exhibit splitting for none of the studied magnetic fields. 

We next consider the situation where the imbalance potential between sublattices is diluted by adding $V_\text{A}$ only on a small percentage $p$ of randomly selected A sites, while keeping the total strength $pV_\text{A}$ fixed for comparison (analogously $-V_\text{A}$ for randomly selected B sites with equal concentration). Note that the random uncorrelated part (characterized by $W$) always remains much stronger then the diluted correlated part. While for $p=100\%$ every A site and every B site are shifted by $V_\text{A}$ and $-V_\text{A}$, respectively, a lower value for $p$ means a random distribution up to an extreme dilution of $0.1\%$ ($pV_\text{A}=0.002$). The corresponding DOS is displayed in Fig. \ref{DOS} (inset) for a magnetic field of 45 Tesla and two values of $W$. It shows that even at low concentrations of $p=1\%$, dilution has negligible effect on the DOS in terms of splitting and peak heights (a long as $pV_\text{A}$ is kept constant).  A further enhancement of the impurities dilution below $p=1\%$ or increase of $W$ ($\geq 0.2$) leads to a disappearance of peaks and splitting signature (Fig. \ref{DOS}, inset). Finally, we note that the splitting does not change the total weight of the DOS, i.e. the integrated DOS is unaffected by the splitting and the peak heights are half of the heights of the initial DOS.

\begin{figure}[htbp]
\begin{center}
\leavevmode
\includegraphics[width=82mm]{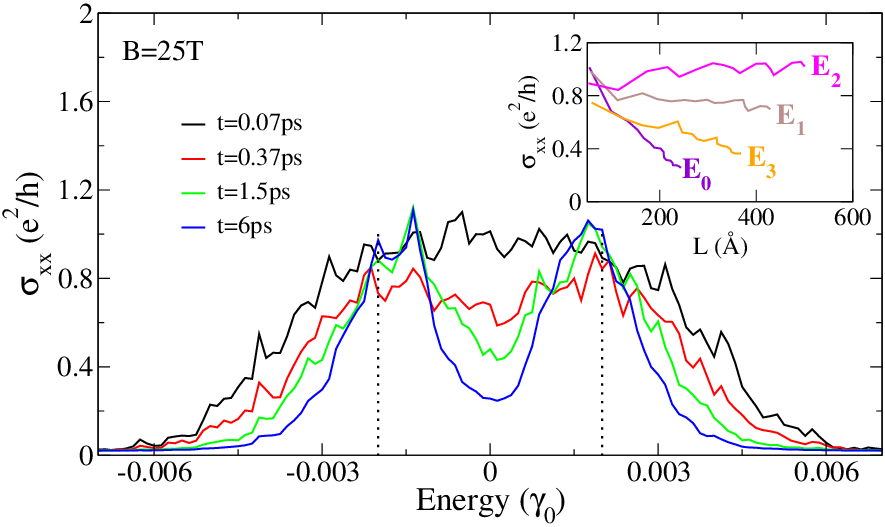}
\caption{(color online) Main frame: Energy dependent conductivity for different times ($p$=0.1$\%$, $pV_\text{A}=$0.002, $W=$0.2). Dotted lines indicate positions $E=\pm pV_\text{A}$ (which, for the case of $p=100\%$ and $W\to$0 indicates the positions of split LL, cf. Fig. \ref{UniversalSigma}). Inset: $\sigma_{xx}$ versus length dependence for selected energies $E_0$=0, $E_k$=0.001$*k$.}
\label{SigmaTime}
\end{center}
\end{figure}

We then scrutinize the time dependence of $\sigma_{xx}$ in the very dilute AB symmetry breaking potential and investigate how robust is the conductivity plateau seen in Fig. \ref{UniversalSigma}. Fig. \ref{SigmaTime} gives the energy-dependence of $\sigma_{xx}$ at 25 Tesla for $p=0.1\%$ ($pV_\text{A}=0.002$, and $W$=0.2). 
One observes a broad feature for the conductivity at small times ($t=20$), which does not show any zero-energy dip.  This is consistent with the corresponding DOS (not shown) which also displays a single maximum (similar to the case of larger $B$ in Fig. \ref{DOS}, inset). The $\sigma_{xx}$ at Dirac point is smaller than $\simeq 1.4e^2/h$ but more importantly displays a strong time dependence indicating the contribution of quantum interferences. Evaluating the quantum conductivity at short times ($t=0.07$ ps) roughly corresponds to introducing an effective cut-off for quantum interferences, thus reducing localization effects. At longer times ($t\geq 0.37$ ps), this broad feature of the profile of $\sigma_{xx}$ is replaced by a double peaks structure, which stems from enhanced contribution of multiple scattering phenomena. Interestingly, the conductivities at the two peak positions (for electron-hole-symmetric points, indicated with dotted lines) are almost identical and marginally affected with time/length scales, which indicates that no localization effects develop at such energies, and corresponding (critical) states remain extended. 

Fig. \ref{SigmaTime} (inset) shows $\sigma_{xx}(L)$ for several typical energies. For $E_{2}$ (vertical dotted lines in Fig. \ref{SigmaTime} (main frame)), $\sigma_{xx}(L,E_2)\sim e^2/h$ and remain length-independent, locating the energy position of the new critical states at the center of the pseudospin-split levels. In contrast, $\sigma_{xx}(L,E_1)$ and $\sigma_{xx}(L,E_3)$ (peak tail) and $\sigma_{xx}(L,E_0)$ (band center) are seen to decay to zero, pinpointing the localization of corresponding states and transition to the insulating Anderson regime.

\begin{figure}[htbp]
\begin{center}
\leavevmode
\includegraphics[width=82mm]{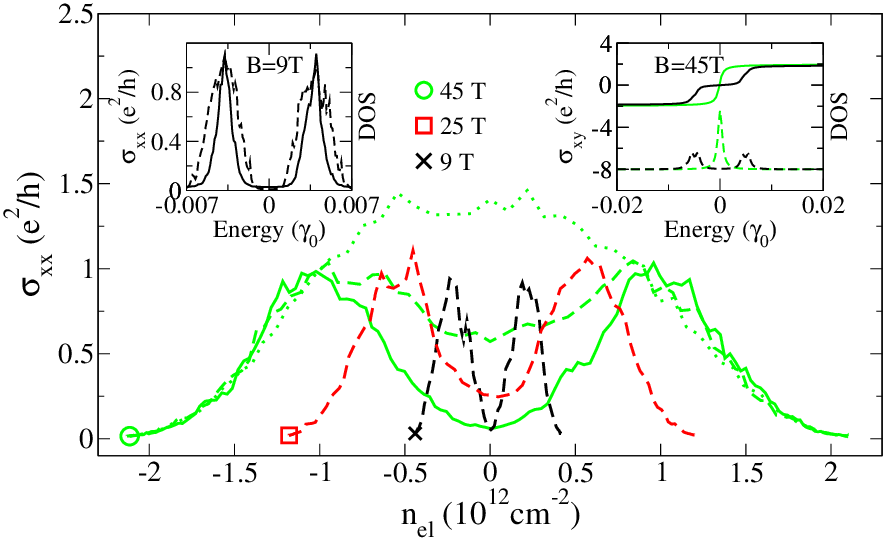}
\caption{(color online) Longitudinal and Hall conductivities with and without sublattice impurity potential. Main frame: $\sigma_{xx}(t$=6ps, $W=0.2$) using $pV_\text{A}$=0.001 (dotted line), $pV_\text{A}$=0.002 (dashed lines), and $pV_\text{A}$=0.004 (solid line). Magnetic fields as indicated. Left inset: $\sigma_{xx}(t$=6ps) (solid line) and DOS (dashed line) for  $pV_\text{A}=0.004$ and $W=0.2$. Right inset: Hall conductivity $\sigma_{xy}(W=0)$ (for $\delta=0.005\gamma_{0},\eta=0.0005\gamma_{0}$, solid lines) and corresponding DOS (dashed lines) for $pV_\text{A}$=0 (green) and $pV_\text{A}$=0.005, $p$=2.5$\%$ (black) at 45 Tesla.}
\label{SigmaChargeDensity}
\end{center}
\end{figure}

The generality of our results is checked by performing a series of calculations for varying magnetic fields and different values for $pV_\text{A}$ which yields neither qualitatively nor quantitatively different results. Our main findings are summarized in Fig.\ref{SigmaChargeDensity} at elapsed computational time $t=6$ps and for $W$=0.2. The maximum value of the doubly peaked $\sigma_{xx}$ turns out to be $B$-independent, which reminds the case of conserved AB symmetry (Fig. \ref{UniversalSigma}). In contrast however, two $\sigma_{xx}$ peaks are clearly observed. Surprisingly, the peak maxima $\sigma_{xx}$ are not half of the maximum obtained in the unsplit case but reduced by a factor of $\simeq 0.7$. This is a clear quantative difference which might be related to the massive/massless character of the Dirac electrons. Earlier works on the 2D electron gas have also debated on the critical value of dissipative conductivity\cite{HuoPRL1993,SchweitzerPRL2005}. Fig.\ref{SigmaChargeDensity} finally shows that $\sigma_{xx}(E=0)\to 0$ while the double-peak height of $\simeq e^2/h$ is robust for different magnetic fields and disorder strength $pV_\text{A}$. 

Finally, we scrutinize the evolution of the Hall conductivity $\sigma_{xy}$ at 45 Tesla for a weak and diluted potential ($pV_\text{A}$=0.005, $p$=2.5$\%$) that breaks A/B sublattice symmetry (inset of Fig. \ref{SigmaChargeDensity}). At the charge neutrality point, the zero-valued plateau $\sigma_{xy}=0$ appears (black solid line) in contrast to the clean case ($pV_\text{A}$=0) where $\sigma_{xy}$ only crosses zero at a single point, when jumping from $-2e^{2}/h$ to $+2e^{2}/h$ (green solid line). The plateau width is here confirmed to be defined by the pseudospin-split states observed in the density of states.  Note that in the case of ultraclean samples electron-electron-interaction effects have been found to also produce additional plateaus in $\sigma_{xy}$\cite{Young2012}.

\textit{Conclusion} - We have reported on robust transport features at the Dirac point for the zero-energy Landau level.  In absence of energy level splitting,  a critical conductivity $\sigma_{xx}\simeq 1.4e^{2}/h$ is obtained for magnetic fields ranging from about 2 to 200 Tesla. When A/B sublattice symmetry is broken by some imbalanced local impurity potential, pseudospin split states are found to convey different critical values $\sigma_{xx}\simeq e^{2}/h$.  A non-dissipative QHE is demonstrated in this model, since $\sigma_{xx}\to 0$ in between pseudospin-split critical states which further dictate the width of the $\sigma_{xy}=0$ quantized plateau. Interestingly, very recent scanning tunneling microscopy experiments on intentionally chemically (nitrogen)-doped or hydrogen-functionalized disordered graphene have revealed the surprising manifestation of some sublattice symmetry breaking mechanism, offering  possibilities for the experimental confirmation of our findings \cite{Zhao2011,Terrones2012,ABSB1,ABSB2}.

\end{document}